%% file: main.tex
\documentclass[journal]{IEEEtran}

\usepackage{color,soul}
\usepackage[ruled,linesnumbered]{algorithm2e}
\usepackage{amsmath,amssymb,amsfonts}
\newtheorem{definition}{Definition}[section]
\usepackage{algorithmic}
\usepackage{amsmath,amssymb,amsfonts}
\usepackage[ruled,linesnumbered]{algorithm2e}
\usepackage[font=small,labelfont=bf]{caption}
\SetKwFor{Foreach}{for each}{do}{endForEach}%
\SetKwFor{For}{for}{do}{endfor for}%
\SetKwIF{uIf}{ElseIf}{Else}{if}{then}{else if}{else}{endIf if}%
\SetKwFor{While}{while}{do}{endWhile while}%
\usepackage{array}
\usepackage{caption}
\usepackage{graphicx}
\usepackage{float}
\usepackage[font=scriptsize,labelfont=bf]{caption}
\usepackage{subcaption}
\usepackage{booktabs}
\usepackage{tabularx}
\usepackage[table]{xcolor}
\usepackage{tabu}
\usepackage{multirow}
\usepackage{tikz}
\usepackage{graphicx}
\usepackage{tikzscale}
\usepackage{pgfplots}
\pgfplotsset{compat=1.17}
\usepackage{makecell}

\makeatletter
\def\@seccntformatinl#1{\csname the#1dis\endcsname\hskip 1em\relax}
\makeatother

\begin{document}

\title{
Privacy-Preserving Ensemble Infused Enhanced Deep Neural Network Framework for Edge Cloud Convergence
}

\author{%
    Veronika~Stephanie,
    Ibrahim~Khalil,
    Mohammad~Saidur~Rahman,
    Mohammed~Atiquzzaman
}


\maketitle

\input{sections/00-abstract}

%
\IEEEpeerreviewmaketitle

\input{sections/01-introduction}
\input{sections/02-related-works}
\input{sections/03-proposed-framework}
\input{sections/04-experimental-results}
\input{sections/05-conclusion}

\bibliographystyle{IEEEtran}
\bibliography{References}

\end{document}

%% file: sections/00-abstract.tex
\begin{abstract}
We propose a privacy-preserving ensemble infused enhanced Deep Neural Network (DNN) based learning framework in this paper for Internet-of-Things (IoT), edge, and cloud convergence in the context of healthcare. In the convergence, edge server is used for both storing IoT produced bioimage and hosting DNN algorithm for local model training. The cloud is used for ensembling local models. The DNN-based training process of a model with a local dataset suffers from low accuracy, which can be improved by the aforementioned convergence and Ensemble Learning. The ensemble learning allows multiple participants to outsource their local model for producing a generalized final model with high accuracy. Nevertheless, Ensemble Learning elevates the risk of leaking sensitive private data from the final model. The proposed framework presents a Differential Privacy-based privacy-preserving DNN with Transfer Learning for a local model generation to ensure minimal loss and higher efficiency at edge server. We conduct several experiments to evaluate the performance of our proposed framework.
\end{abstract}


%% file: sections/01-introduction.tex
\section{Introduction}

\IEEEPARstart{T}{he} massive improvement in Internet-of-Things (IoT) technology has enabled rapid data collection in different applications, including healthcare. For example, Alexapath has developed an IoT-enabled microscope for instant collection of microscopic images, which can be shared with specialists working remotely \cite{auguste2019validation}. Giving another example, an IoT device can capture lung images by passing a small amount of current on cross areas of human lung\footnote{https://buildforcovid19.io/lung-imaging-iot-for-remote-monitoring-in-isolation/}. The captured data can later be used for diagnosis and research work by the health practitioners by leveraging Artificial Intelligence (AI) techniques such as Deep Learning (DL)\cite{9151167}. Nevertheless, IoT devices are resource-constrained and cannot offer data storage and DL tasks alone due to the computational power requirements of AI tasks. 

Cloud is a popular platform for traditional IoT data storage and DL-based machine learning in both industry and academia. However, the cloud is not suitable for realtime data analysis services due to the high bandwidth requirement and network latency \cite{bugshan2021privacy}. Edge computing technology is getting attention from machine learning practitioners and researchers due to several reasons. \textit{First}, edge devices can be integrated with IoT devices for rapid data collection and to store data in private edge data servers. \textit{Second}, a part of cloud DL tasks can be offloaded in the edge servers and executed with private data to reduce bandwidth requirements. \textit{Third}, edge devices enable the convergence of IoT, edge, cloud, and AI to solve machine learning tasks at the close proximity of data source and offer realtime services. By leveraging the aforementioned convergence, a healthcare service provider (e.g., hospital) can deploy several IoT devices and edge devices to collect bioimages and store them, respectively. In addition, hospitals can deploy edge devices to host DL algorithms to train models for data analysis based on the collected bioimages. 

The accuracy of the DL approach is a necessary requirement that depends on the variety of used models. A hospital may apply a model which may not be enough to achieve high prediction accuracy during data analysis. Hence, the hospital may need to collaborate with other hospitals to improve the accuracy of prediction. Moreover, a dataset of a single hospital may be homogeneous, which leads to model overfitting and causes significant utility loss in a DL model \cite{kaissis2021end}. Hence, collaborating datasets with other hospitals should increase the dataset volume and help improve the DL model accuracy. This large-scale dataset is often obtained from multi-institutional or multi-national data accumulation, and voluntary data sharing \cite{kaissis2021end}.

Nevertheless, collaborating datasets with other hospitals introduces a privacy risk as the dataset contains sensitive information about the patient. In a centralized deep learning model training process, it is common for medical institutions to anonymize or pseudonymize patients' data before sending it to public analysis and model training sites. However, it is proven that anonymization is insufficient to protect against re-identification attack \cite{branson2020evaluating}. Moreover, once the anonymized medical data are transmitted to a public site, the data cannot be easily revoked or augmented \cite{tan2021towards}. 

An ensemble learning method is a collective machine learning approach to obtain better prediction performance by strategically combining multiple learning models. The ensemble learning approach gives high accuracy without sufficient data representation \cite{krawczyk2018online}. The training process of DNN is training a loss function to find out a set of weights that are considered suitable for a given problem. The loss surface in a complex network is more chaotic with many local optimal solutions \cite{10.5555/3327345.3327535}. Ensemble learning effectively utilizes these multiple local optimal solutions to improve the accuracy of the prediction significantly \cite{liu2018learning}. Therefore, ensemble learning is a suitable mechanism for convergence IoT, edge, cloud, and AI.

\subsection{Problem Statement}
\begin{figure}[ht!]
\centering
\includegraphics[width=1\linewidth]{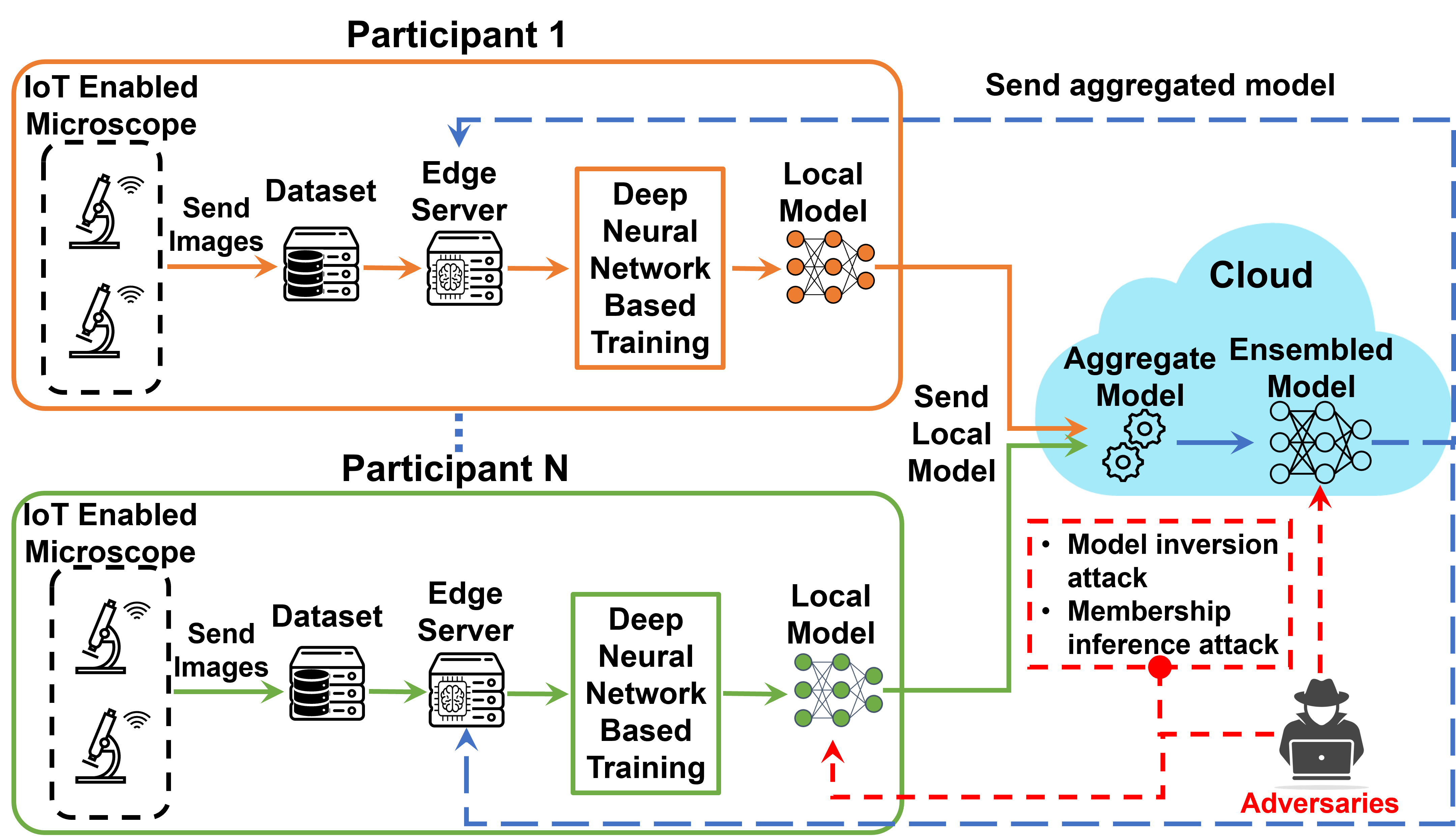}
\caption{Centralized collaborative learning without privacy-preservation method}
\label{fig: probStatement}
\end{figure}

Although ensemble learning improves the model accuracy, this approach alone is not sufficient to provide users privacy \cite{truong2021privacy}. To describe the problem in ensemble learning-based collaborative deep learning model, we choose a healthcare scenario as shown in Fig. \ref{fig: probStatement}. We assume that several hospitals, called participants, collect patient lung images via IoT devices and store them in private edge data servers. Hospitals also own private edge servers that host DL algorithms. A private edge server of a hospital learns from local data to generate a local training model. All private edge servers outsource local models to a centralized public cloud server that hosts the ensemble algorithm. The public cloud server aggregates received models using an ensemble method to generate a final model shared with all hospitals for their data analysis.  

As shown in Fig. \ref{fig: probStatement}, an adversary can perform several attacks on shared local models and the final models to leak sensitive information. Authors in \cite{fredrikson2015model} exhibit a successful model inversion attack that utilizes shared local model parameters on collaborative learning setting to reconstruct most of the data used for training. \cite{shokri2017membership} developed a membership inference attack that can determine whether a record was used as part of the machine learning model's training. Hence, existing studies have tried to combine the distributed learning models with some privacy-preserving methods such as Secure Multi-party Computation (SMPC) \cite{ma2018deep, sotthiwat2021partially}, Differential Privacy (DP) \cite{xiang2018collaborative, wei2020federated} and Homomorphic Encryption (HE) \cite{cheon2018ensemble} to enhance the system's privacy. However, the integration of the privacy preservation method and distributed DL system introduces another issue. For example, in distributed learning using DP, adding too much noise yield poor performance of the resulting model. On the other hand, in HE and SMPC integrated distributed DL systems, massive computation and communication overhead are their profound issues. Furthermore, in most distributed DL systems, as shown in Figure \ref{fig: probStatement}, additional communication overhead is usually found due to constant local model exchange with the server to enhance the performance of the public model.

\subsection{Contributions}

In this paper, we propose a privacy-preserving ensemble infused enhanced Deep Neural Network (DNN) based learning framework for IoT, edge and cloud convergence. Initially, we propose a privacy-preserving architecture for the edge and cloud-based ensemble-assisted DNN framework. In the proposed architecture, multiple participants train their own models individually at their private edge servers based on the local dataset. Next, the local training model generation process is made privacy-preserving with Differential Privacy to prevent privacy leakage. We use Stochastic Gradient Descent (SGD) based mechanism in Differential Privacy to add noise to the training model parameters. As applying Differential Privacy results in a significant loss in the model, we apply Transfer Learning\cite{8689016} to mitigate the loss. We assume that a trusted third party generates an initial model using a Convolutional Neural Network (CNN) with a public dataset before beginning the local model generation. An edge server gets the initial model from the trusted third party and transfers the knowledge to repair the loss. Applying transfer learning also improves efficiency and reduces the computational load of the edge server. Finally, an ensemble-based collective mechanism is developed to generate a final model. The ensemble process is performed and the final model is distributed by the cloud.

Our contributions are summarized as follows:
\begin{itemize}
  \item A framework is proposed for Deep Neural Network (DNN) based learning to ensure high learning accuracy in IoT, edge, and cloud convergence.
  \item Our framework leverages ensemble learning concept to generate a generalized final model which is robust and ensures good accuracy in the context of IoT, edge, and cloud convergence.
  \item Differential Privacy-based privacy-preserving technique is used with Transfer Learning during local model generation to protect from privacy attacks with higher efficiency and reduced computational load at edge servers. 
\end{itemize}

\subsection{Organization}
The remainder of this paper is organized as follows. Section \ref{sec:related_work} discusses some of the closely relevant works. Some preliminary topics are presented in Section \ref{sec:pre}. The system architecture of the proposed framework is presented in Section \ref{sec:archi}. The methodology used in the proposed framework is described in Section \ref{sec:method}. Experimental results and performance evaluation are shown in Section \ref{sec:exp}. Finally, Section \ref{sec:con} concludes this paper.

%% file: sections/02-related-works.tex
\section{Related Work}\label{sec:related_work}

To preserve privacy in collaborative learning scheme, recent studies tried to integrate it with other privacy-preserving methodologies, such as Multi-Party Computation (MPC), Homomorphic Encryption (HE), and Differential Privacy (DP).

\cite{phong2019privacy} proposed a privacy-preserving deep learning system via weight transmission. In preserving the model privacy, participants share symmetric keys to keep the model secret from the server. The server acts as a transfer station for the local model to be distributed and trained. The fact that the training model needs to be updated and trained in a sequence is not efficient.

A method proposed by \cite{9066956} takes advantage of a centralized and distributed training scheme to achieve efficiency and reduce computational cost. The proposed model employs differential privacy, blinding, and HE techniques to ensure the model's ability to resist collusion attacks, model attacks, and data attacks.

Another DP based privacy-preserving collaborative learning introduced by \cite{shokri2015privacy}. To minimize information leakage from the shared model, they only selected some gradients over a certain threshold to be transmitted. However, the proposed method is claimed to provide moderate privacy as a subset of the parameters is shared with other participants for each training iteration \cite{ma2018deep}. 

In tackling the issue, \cite{ma2018deep} proposed a collaborative deep learning scheme, whose privacy preservation method is based on secure MPC. The proposed model shows that the scheme is able to protect the local dataset and learning model from the cloud server. \cite{9235504} also proposed a secure MPC based collaborative learning model which is resistant to generative adversarial network based attack by ensuring that participants are isolated from the model parameters. However, due to its high cost in calculating complex functions, its application in a privacy-preserving deep learning environment may not be suitable. Focusing on reducing computational cost, \cite{sotthiwat2021partially} proposed partial model encryption using MPC in the distributed deep learning system. Although the computational cost is significantly reduced, the underlying high communication cost problem of MPC can still be found where synchronous updates are required through all participants for each training iteration. 

Aside from MPC, a recent study in privacy-preserving collaborative deep learning using HE is also proven to provide privacy guarantees against honest-but-curious servers and parameter leakage. For example, \cite{phong2018privacy} introduces a privacy-preserving deep learning model using an additive HE scheme. The proposed scheme encrypts local model's gradients before being sent to the server. The model is proven to be secure against curious servers at the cost of increased communication between the learning participants and the cloud server.

The aforementioned method above tries to collaboratively enhance public model performance by updating model parameters in a privacy-preserving manner. On the other hand, one of the state-of-the-art approaches, Private Aggregation of Teacher Ensembles (PATE) proposed by \cite{papernot2018scalable}, is a DP based method that allows us to predict an unlabeled public data in a privacy-preserving manner to train another learning model. One method that is closely related to our work is proposed by \cite{8689016}, which is a transfer learning based PATE ensemble learning method. This method tries to transfer the knowledge of existing public data to each of the ensemble learning teacher models. This allows the proposed method to increase the performance of their model in predicting unlabeled data. However, when a different participant owns a teacher model in a collaborative learning scheme, the proposed method may cause a privacy leak. This is because each teacher model is made visible to other participants. Hence, the model parameters can be used to do a membership inference attack or model inversion attack on a specific participant.

From the discussed methods above, we note that each privacy preservation method has trade-offs between accuracy, computation cost, and communication cost. HE and MPC based method provides better accuracy and privacy at the cost of increased communication and computational cost. On the other hand, DP based methods are more efficient in terms of their communication and computational cost in return for their performance loss. In addition to that, most of the privacy-preserving distributed learning methods mentioned above consider periodical local model updates from the local devices and then send the aggregated model back to all devices. This results in substantial communication overhead \cite{mohammadi2021differential}. While considering both aspects, we propose a high-performing ensemble distributed learning system with knowledge transfer based on differential privacy, which does not require a continuous local model update.

%% file: sections/03-proposed-framework.tex
\section{Preliminaries}\label{sec:pre}
In this section, we present the background of DL, ensemble learning, distributed learning, and DP that serve as the fundamentals in this article.

\subsection{Deep Learning}
DL is a branch of machine learning that allows us to gain a high level of abstractions on a set of data. Essentially, DL neural network is composed of more than three layers. These layers are proposed to imitate the human brain. This allows DL to learn from a large amount of data, hence its success on complex tasks such as computer vision, image classification, and language processing. Typical process in deep learning usually involves forward propagation and backpropagation. Forward propagation allows us to retrieve output value from the given input data, while backpropagation updates the parameter of the neural network model. To update the weight parameters, backpropagation has an optimizer which role is to calculate loss and update the model parameters to reduce the loss. Generally, a loss function can be denoted as $l(y, \hat{y})$, where $y$ is the true label and $\hat{y}$ is the prediction. In Stochastic Gradient Decent (SGD), we can compute the updated parameter as follows.
$$\Theta_{k+1}= \Theta_{k}-\eta \cdot \frac{1}{N} \sum_{i=1}^N \nabla_{\Theta_{k}}l(f(x_{i}), y_{i})$$
Here $\Theta_{k}$ is the parameter of the current step $k$, $\eta$ is the learning rate, $N$ is the number of samples within a batch, $\nabla$ is used to refer to the derivative with respect to every parameter, and $l(f(x_{i}), y_{i})$ is our loss function, which takes prediction of an input data $x_{i}$ from a prediction function $f()$ and a true label of the input data $y_{i}$ as the inputs.

\subsection{Ensemble Learning}
Ensemble learning is introduced as a method that combines multiple learning models, whose primary goal is to improve the capability of its base models $\mathcal{M}$. Ensemble learning can be classified into three classes. They are bagging, boosting, and stacking. In this particular paper, we focus on the use of the ensemble averaging method, which is a Bootstrap Aggregation or bagging based ensemble learning introduced in \cite{hashem1997optimal}. In Deep Neural Network (DNN), only one model is usually kept for training and predicting a dataset. However, in ensemble averaging, several neural network models are kept, and the prediction obtained by each learning model are aggregated to reduce the base model bias and variance error. Generally, ensemble averaging can be calculated using Equation \ref{eq: weightedEns}.
\begin{equation}
     {\tilde {P}}(x ,\mathbf {\alpha } )=\sum _{i=1}^{k}\alpha _{i}P_{i}(x)
    \label{eq: weightedEns}
\end{equation}
Here, $\tilde {P}()$ represents the predictions of the ensemble model, $x$ is input data, $\alpha$ is a list of weights, where $\alpha_{i} \in \alpha$ is a weight assigned to a particular base model $\mathcal{M}_i \in \mathcal{M}$, $k$ is the number of participants, and $P_i(x)$ is the resulting prediction probabilities of input data $x$ from base model $\mathcal{M}_i$. A raw average in ensemble averaging can be achieved by replacing the value of $\alpha_{i} \in \alpha$ with the value of one over the total number of $\mathcal{M}$.

\subsection{Transfer Learning}
Transfer learning can be seen as a mechanism that allows a system to utilize a learned knowledge from one task to another task \cite{chris2017multi}. This can help address limitations in medical fields where healthcare image data can hardly be shared for DL model training.

Transfer learning comprises two concepts: domain $D$ and learning task $T$. Formally, a domain can be represented as $D = \{\mathcal{X}, P(X)\}$. Here, $\mathcal{X}$ represents the feature space, and $P(X)$ is the marginal probability distribution of sample $X$ in $\mathcal{X}$. A task can be denoted as $T = \{\mathcal{Y}, P(Y|X)\}$. Here, $\mathcal{Y}$ is the label space, and $P(Y|X)$ is a predictive probability function, which predicts the conditional probability of $Y \in \mathcal{Y}$ given $X \in \mathcal{X}$.

Given a target task $\mathcal{T}_t$ and a target domain $\mathcal{D}_t$, we can transfer the knowledge from a source domain $\mathcal{D}_s$ with a source task $\mathcal{T}_s$ to improve the performance of the predictive function in $\mathcal{T}_t$, where $\mathcal{D}_t \neq
 \mathcal{D}_s$ and/or $\mathcal{T}_t \neq
 \mathcal{T}_s$.

A typical transfer learning can be done by transferring the learned parameters of a well-trained DL model on a large dataset (e.g., ImageNet) to the $\mathcal{D}_t$. In this project, we focus on the use of transfer learning in a Deep Neural Network (DNN), where the source and target have the same domain and task.

\subsection{Differential Privacy}
DP is a mechanism that aims to minimize the risk of privacy breaches in a particular database. The definition of differential privacy can be formalized as follows.

\begin{definition} 
A randomized mechanism ($M$) provides ($\varepsilon$, $\delta$)-differential privacy if any datasets $D$ and $D'$ that differ at most one element, and for any subset $S \subseteq Range(M)$, where $range(M)$ represent the range of possible outcomes produced by $M$,
\end{definition}
\begin{equation}
    P(M(D) \in S) \leq e^\varepsilon \times P(M(D') \in S) + \delta
\label{eq: DiffPrivDefinition}
\end{equation}

As given in Equation \ref{eq: DiffPrivDefinition}, the $\varepsilon$ is privacy metric loss which provides an insight into the loss of privacy in the corresponding differentially private algorithm. Initially, the original differential privacy was proposed by \cite{dwork2008differential} as $\varepsilon$-differential privacy. However, to loosen the definition of $\varepsilon$-differential privacy, $\delta$ was introduced. $\delta$ is defined as the probability of information leakage accident. This $\delta$ is expected to be smaller than $\frac{1}{|D|}$ where $D$ is the data size within the database.

\begin{figure*}[ht!]
\centering
\includegraphics[width=0.8\linewidth]{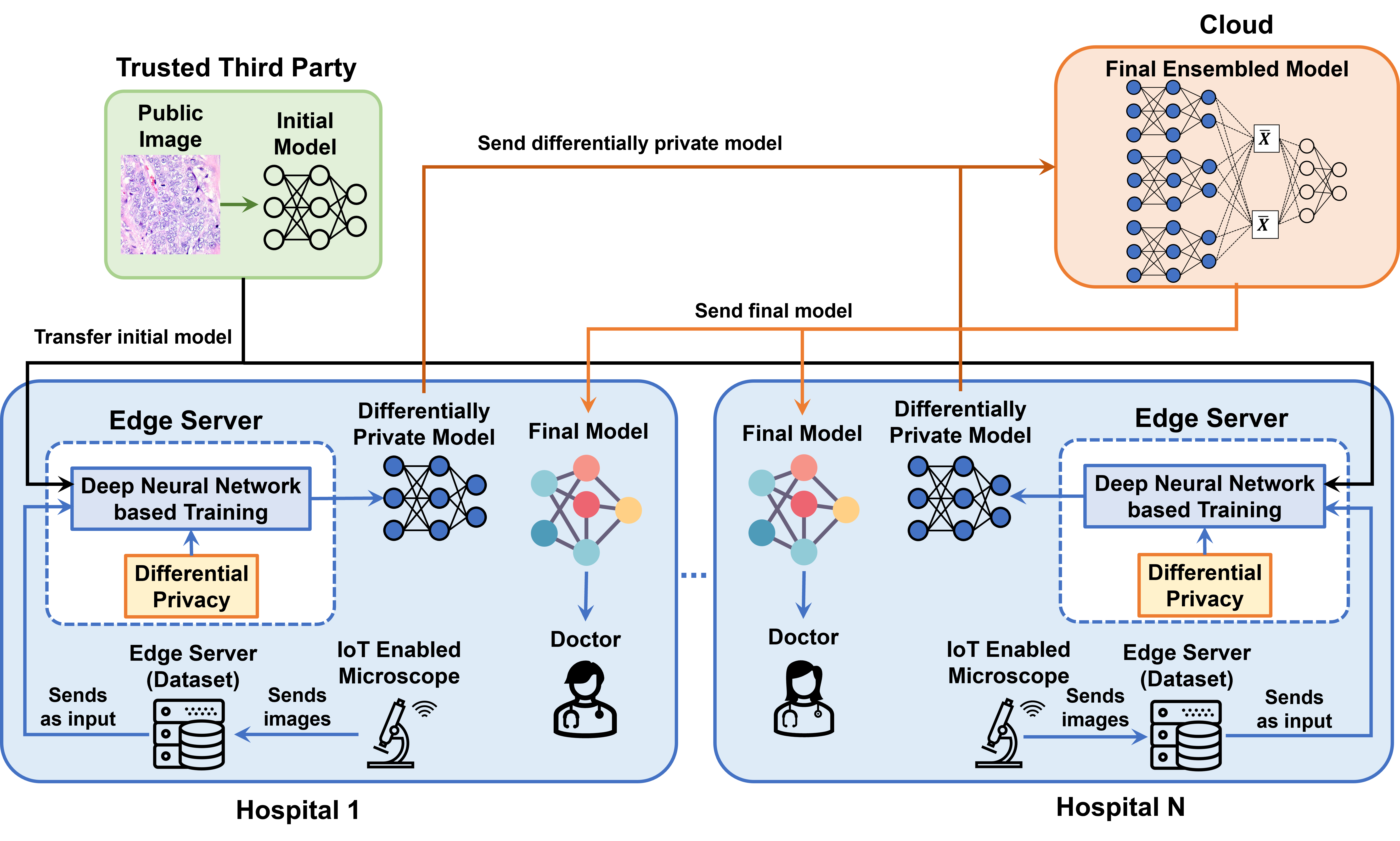}
\caption{Overview of the proposed architecture}
\label{fig: ArchOverview}
\end{figure*}

\section{Proposed System architecture}\label{sec:archi}
Figure \ref{fig: ArchOverview} shows the architecture overview of the proposed method. We assume that there are $n$ hospitals, denoted as a set $\{H_1, H_2, \hdots, H_n\}$, in the proposed systems. Each hospital $H_i (1 \leq i \leq n)$ comprises of several entities, namely, IoT devices, clients, edge nodes, and cloud. The roles of each component are described below:

\begin{itemize}
  \item \textit{IoT devices:} In the proposed framework, IoT devices are owned by a hospital $H_i$. Each IoT device is integrated into a medical device and acts as a data source. IoT devices of $H_i$ capture medical images and store them in a local dataset. For $n$ hospitals, there are $n$ local datasets.
  \item \textit{Edge Data Server:} An edge data server $E_{Di}$ is an edge device owned by hospital $H_i$ that stores local dataset generated by IoT devices of $H_i$. Data in $E_{Di}$ is considered private and cannot be shared with other hospitals to ensure privacy. For the sake of simplicity, we assume that each $H_i$ owns a single edge data server $E_{Di}$.
  Users from the same hospital are connected to the same edge data server, while users from different hospitals must be connected to different edge data servers. The data in the edge data server $E_{Di}$ is used to train a model for the hospital $H_i$ privately.
  \item \textit{Private Edge Server:} A private edge server $E_{Si}$ is an entity that is locally owned by a hospital $H_i$. The private edge server $E_{Si}$ uses local dataset in $E_{Di}$ to train an initial model $M_I$ locally. $E_{Si}$ gets an $M_I$ from a trusted source called \textit{trusted third party}. $M_I$ is distributed among private edge servers $\{E_{S1}, E_{S2}, \hdots, E_{Sn}\}$ of all hospitals to create their respective privacy-preserving training models.
  \item \textit{Trusted Third Party:} In our proposed framework, the trusted third party (TTP) is a secure cloud. TTP leverages a public dataset to train a model (i.e., $M_I$) using CNN based deep neural network. We discuss the detailed process later in this section. 
  \item \textit{Cloud:} The cloud is a public entity that collects all locally trained models from all hospitals. The set $M$ of locally trained models are represented as set $\{M_1, M_2, \hdots, M_n\}$. The locally trained models are ensembled together to generate an aggregated model ${M}_{\Sigma}$, which is then sent to all private edge servers for updating their respective local models.
\end{itemize}

\section{Methodology}\label{sec:method}
\vspace{-1mm}In this section, we discuss our proposed framework for privacy-preserving high-performing ensemble assisted DL with Transfer learning in edge cloud consortium. Our proposed architecture has three major steps. The first step includes generating an initial model from a public dataset. The second step is generating a locally trained model based on a local dataset while preserving privacy. The final step ensembles all local training models to obtain an aggregated model. Each of the steps are discussed below.

\vspace{-5mm}
\begin{table}
\fontsize{10}{14}\selectfont
\begin{center}
\caption{Notations}
\scalebox{0.8}{
{
\begin{tabularx}{\linewidth}{
    >{\hsize=.3\hsize}X
    >{\hsize=.6\hsize}X
}
\toprule
    $H_{i}$ & Hospitals\\
    $E_{Di}$ & Edge Data Server\\
    $E_{Si}$ & Private Edge Server\\
    $M_{I}$ & Initial Model\\
    $M_{n}$ & Locally Trained Model\\
    $D_{pub}$ & Public Dataset\\
    $ep$ & Epochs\\
    $b$ & Batch Size\\
    $D_{label}$ & A Set of Labeled Data Partitions\\
    $\overline{y}$ & Prediction on Input Data\\
    $\mathcal{L}$ & Computed Loss\\
    $grad$ & Local Computed Gradient\\
    $DB_{Li}$ & Local Dataset owned by $H_{i}$\\
    $M_{Pi}$ & Locally Trained Private Model\\
    $nm$ & Noise Multiplier\\
    $nc$ & Clipping threshold\\
    $mb$ & Mini-batch Size\\
    $l$ & Layer Partition\\
    $L$ & Layers within CNN Model\\
    $\epsilon$ & Privacy Budget\\
    $\delta$ & Probability of Information Leakage Accident\\
    $M_{P}$ & Local Private Models\\
    $M_{E}$ & Ensembled model\\
\bottomrule
\end{tabularx}
}
}
\end{center}
\end{table}

\subsection{Initial Training Model Generation}
The first step of the proposed framework is the generation of \textit{initial model} $M_{I}$ by a TTP. TTP takes a public dataset $D_{pub}$ as input and applies Convolutional Neural Network (CNN)\cite{lecun1998gradient} to train $M_{I}$. A typical CNN consists of three types of operation layers: \textit{convolutional layer} (CONV), \textit{pooling layer} (POOL), and flattening (FLAT) and fully connected layer (FC). The CONV layer consists of multiple sub-layers that are used for feature extraction. The POOL layer acts as the merging layer. In initial model generation, we use \textit{max pooling}. The FLAT layer formats the extracted features to forward to the FC layer.  

\begin{figure}[ht!]
\centering
\includegraphics[width=1\linewidth]{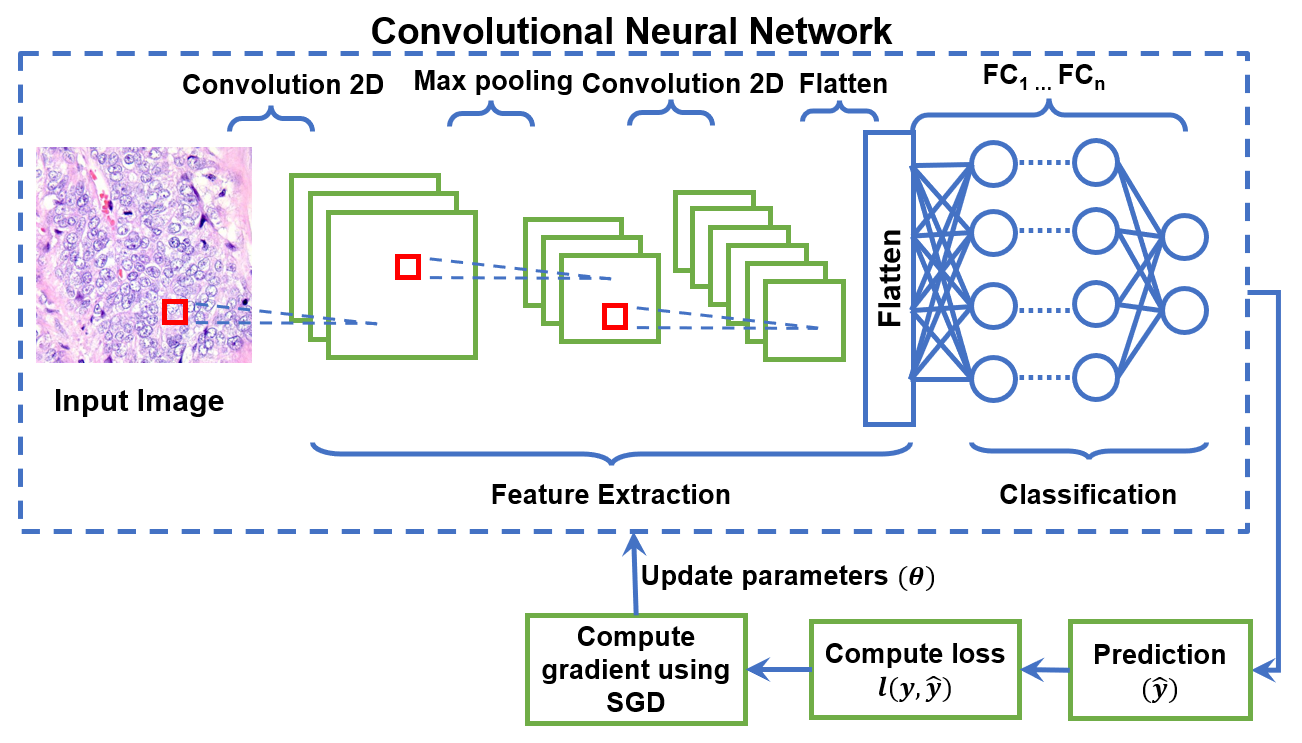}
\caption{Generation process of Initial Model by Trusted-Third Party}
\label{fig: init_model}
\end{figure}

An overview of the initial model generation process is illustrated in Fig. \ref{fig: init_model}. The process is summarized in Algorithm \ref{alg: initModel}. In the initial model generation process, we use the Lung Cancer \cite{borkowski2019lung} public dataset that contains 15K 2D-images.  
Initial model training is performed by a trusted third party. TTP creates $M_{I}$ with random parameter $\theta$. During this process, TTP initializes the other model training parameters, such as the number of training epochs $ep$ and batch size $b$. In each epoch, TTP uses an optimizer to update parameters and a loss function to calculate how well the model is performing. We use the \textit{Categorical Crossentropy} function \cite{zhang2018generalized} as the loss function and the \textit{Stochastic Gradient Decent (SGD)} \cite{kiefer1952stochastic} as the optimizer, which formula can be seen in Equation \ref{eq: cce} and Equation \ref{eq: sgd} respectively.

\begin{equation}
    l(y, \hat{y}) = -\frac{1}{N}\sum^{N}_{i=0}\sum^{J}_{j=0} y_j \cdot log (\hat{y}_j) + (1-y_j) \cdot log(1-\hat{y_j}).
\label{eq: cce}
\end{equation}

Here, $N$ is the number of data, $j$ is the number of classes, $y$ is a vector representing true label, and $\hat{y}$ is a vector representing the probability of class prediction.

\begin{equation}
    \theta_{t+1} = \theta_{t} - \eta_{t} \cdot \frac{1}{N} \sum_{i=1}^N \nabla_{\theta_{t}}l(y_{i}, \hat{y}_i),
\label{eq: sgd}
\end{equation}
where, $t$ is the current step, $\eta$ is the learning rate, and  $\nabla$ is the derivative with respect to every parameter.

In the CNN-based training process, we first randomly sample data from $D_{pub}$ using a function called $trainLoader()$ according to the batch size $b$ and results in multiple partition with labeled data $D_{label}$. An element in $D_{label}$, represented as $D^{j}_{label}$, is a pair $\{input_{j}, label_{j}\}$. Here, $input_j$ is an image with $label_j$. Each partition holds $b$ number of data which is feed into the initial model $M_{I}$. Next, current prediction $\overline{y}$ is computed for $D^{j}_{label}$ using a function $computePrediction()$. The prediction $\overline{y}$ is then used to compute the loss $\mathcal{L}$ of the model along with the true label of the input data using a function $computeLoss()$. Further, we compute the gradient $grad$ from the loss $\mathcal{L}$ using a function $computeGradient()$. Finally, the SGD parameters of the initial model $M_I$ are updated using the function $updateSGDParam()$ with the current $grad$. Once the model is finalized, $M_I$ is ready to be collected by any participants (i.e., hospitals) from TTP. 

\SetAlFnt{\small}
\begin{algorithm}[ht!]
\SetAlgoNoLine
\caption{Initial model training}
\label{alg: initModel}
\KwIn
{
    \begin{minipage}[t]{10cm}%
     \strut
        Public dataset, $D_{pub}$ 
     \strut
    \end{minipage}%
}
\KwOut{
    \begin{minipage}[t]{8cm}%
     \strut
        Initial public model, $M_{I}$ 
     \strut
    \end{minipage}%
}

    \textbf{Initialization:}
    
    Initial CNN model, $M_{I} = \emptyset$
    
    number of epochs, $ep$
    
    Batch size, $b$

    \textbf{begin}
    
    \Foreach{$i \leq ep$}
        {
            $D_{label} \leftarrow trainLoader(D_{pub}, b)$
            
            \Foreach{$\{D^{j}_{label}\} \in D_{label}$}{
                $\overline{y} \leftarrow D^{j}_{label}.computePrediction()$
                
                $\mathcal{L} \leftarrow D^{j}_{label}.computeLoss(\overline{y})$
                
                $grad \leftarrow D^{j}_{label}.computeGradient(\mathcal{L})$
                
                $M_I.updateSGDParam(grad)$
            }
        }

    \textbf{return} $M_I$
    
    \textbf{end}

\end{algorithm}

\subsection{Generating a Locally Trained Model with Privacy}

\begin{figure}[ht!]
\centering
\includegraphics[width=1.0\linewidth]{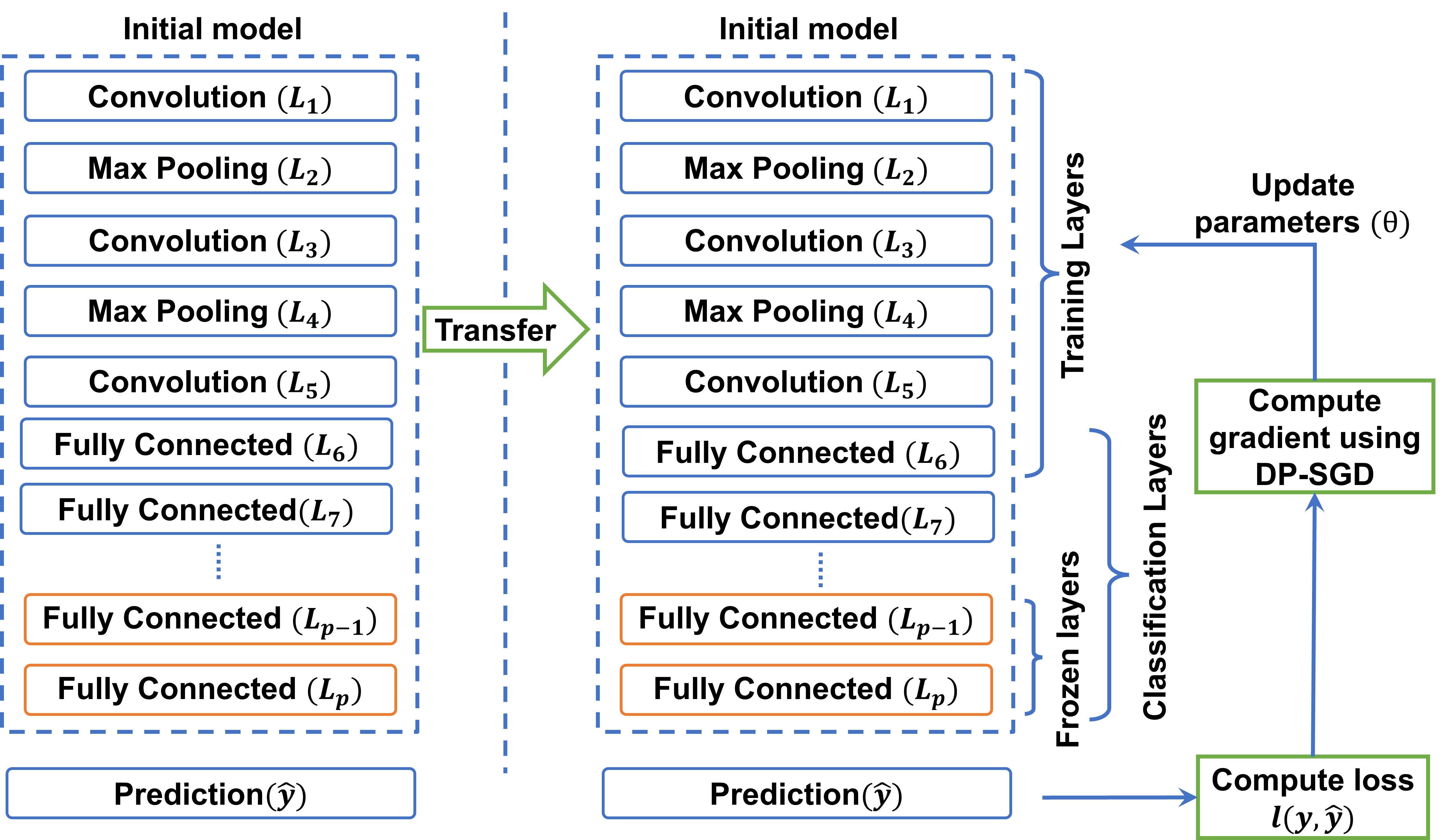}
\caption{Local model transfer and private model training}
\label{fig:localmodelgen}
\end{figure}

In this step, each participant (i.e., hospital) $H_i$ generates a local training model $M_{Li}$ at the private edge server $E_{Si}$ from their local dataset $DB_{Li}$. As $DB_{Li}$ contains sensitive information, $E_{Si}$ applies Differential Privacy (DP)-based privacy-preserving mechanism. An $E_{Si}$ collects an initial model $M_I$ from TTP and applies CNN with Transfer Learning (TL) approach to generate the local model. An overview of the process is illustrated in Fig. \ref{fig:localmodelgen}.

In our proposed training process (see Algorithm \ref{alg: privModel}), we assume that there are $p$ layers that are denoted as $\{L_{1},L_{2}, \hdots, L_{p-1},L_{p}\}$. Initially, $E_{Si}$ freezes the last two layers (i.e., $L_{p-1}$ and $L_{p}$ ) of the initial model $M_{I}$ and executes first $p-2$ steps of CNN. We assume that the local model $M_{Li}$ has a model parameter $\theta_i$. To minimize the privacy leakage of $\theta_i$, a differentially private SGD is applied which is named as DP-SGD \cite{abadi2016deep} in the CNN layers. The \textit{parameter update} process in DP-SGD is similar to original SGD. Nevertheless, noise is added with the parameters to ensure the privacy of $\theta_{i}$. Let, $X$ is the set of private data such that $X \subseteq DB_{Li}$. At first, a set $\overline{X}$ of random data points of size $m$ is selected from $X$. The set of data points $\overline{X} = \{x_{1}, x_{2}, \hdots, x_{m}\}$. Next, gradients are computed for each $x_k \in \overline{X} (1 \leq k \leq m)$ as follows:
\begin{equation}
    g_{t}[x_k] = \nabla_{\theta_{k}}l(f(x_{k}), y_{k}),
\end{equation}
where, $t$ is the current step, $\theta_{k}$ is the current state of the model parameter, $\nabla$ is used to refer to the derivative with respect to every parameter, $f(x_{k})$ is the model prediction with respect to input $x_{k}$, $y_{k}$ is the true label of input $x_{k}$, and $l()$ is the loss function.
Finally, the gradients are used to update the model parameters. 

To apply differential privacy during the local model training, DP-SGD uses a few additional steps after the gradient computation. The additional steps includes \textit{gradient norm clipping} and \textit{noise addition}. The gradient norm clipping limits how each individual training point is sampled in a mini-batch and influences the resulting gradient computation. In DP-SGD, the gradient norm clipping can be calculated as follows:
\begin{equation}
    g_{t}[x_k] = \frac{g_{t}[x_k]}{max[1, ||g_{t}[x_k]||/nc]},
\end{equation}
where $nc$ is the level 2 (L2) norm clip threshold. This threshold is to ensure $L_2$-sensitivity since the privacy guarantee in Gaussian mechanism requires that the noise vector standard deviation of each coordinate to scale linearly with the $L_2$-sensitivity of gradient estimate $g_{t}[x_k]$
\cite{chen2020understanding}.  
Noise addition step adds a noise to the clipped gradient to provide privacy to the model. DP-SGD uses Gaussian noise mechanism to calculate the noise, which can be shown using the following equation:
\begin{equation}
  \xi_{t,i} \sim Norm_{\xi_{t,i}}[0,nm^{2} nc^{2} mb],  
\end{equation}
\begin{equation}
   g_{t}[x_k] = \frac{1}{L} (\sum g_{t}[x_k] + [0,nm^{2} nc^{2} mb])
\end{equation}
where $nm$ is a noise multiplier value, and $mb$ is mini-batch size. 

The last two layers of the CNN model produced by this training process are then removed to produce a differentially private model $M_{Pi}$ of the private edge server $E_{Si}$ for the hospital $H_i$. This allows the output of each private model to be averaged together and produce an ensembled output to be fed as an input to the last two layers of the initial model $M_{I}$. $E_{Si}$ sends $M_{Pi}$ to cloud for ensemble.

\begin{algorithm}[ht!]
\SetAlgoNoLine
\caption{Generating local private model $M_{Pi}$}
\label{alg: privModel}
\KwIn
{
    \begin{minipage}[t]{10cm}%
     \strut
        Initial public model $M_{I}$, Local data $DB_{Li}$ of $E_{Si}$
     \strut
  \end{minipage}%
}
\KwOut{
    \begin{minipage}[t]{8cm}%
     \strut
        Local private model of $E_{Si}$, $M_{Pi}$
     \strut
    \end{minipage}%
}
{
    
        \textbf{Initialization:}
        
        Noise multiplier, $nm$ \\
        Clipping threshold, $nc$ \\
        Mini-batch size,  $mb$ \\
        layer partition,  $l$ \\
        Number of epochs, $ep$\\
        Set of layers, $L = \{L_1, L_2, \hdots, L_p\}$
    
    \textbf{begin}
    
    \Foreach{$e \in ep$}{
        
        $\overline{X} \leftarrow trainLoader(DB_{Li}, mb)$
        
        \Foreach{$L_{j} > (p-2)$}{
            $M_{I}.freeze(L_{j})$
        }
        
        \Foreach{$x_{i} \in \overline{X}$}{
            $\overline{y} \leftarrow computePrediction(x_{i})$
            
            $\mathcal{L} \leftarrow l(\overline{y}, M_{I}.labels)$
            
            $grad \leftarrow computeGradient(\mathcal{L})$
            
            $M_{Pi}.updateSGDParameter(grad)$
        }
    }
    
    \textbf{return} $M_{Pi}$
    
    \textbf{end}
}

\end{algorithm}

\begin{figure}[ht!]
\centering
\includegraphics[width=1.0\linewidth]{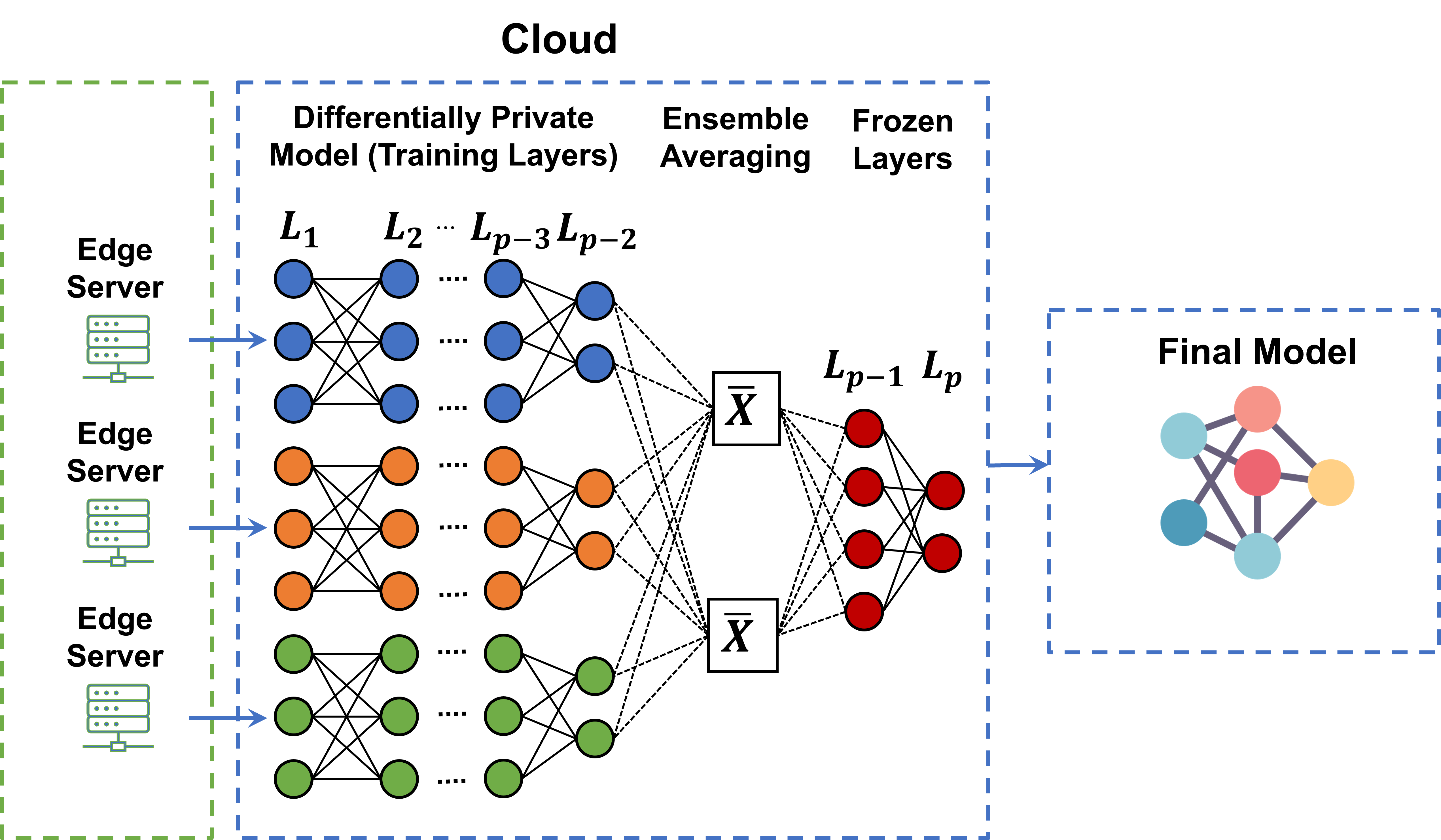}
\caption{Ensemble assisted aggregated model generation}
\label{fig:ensemble}
\end{figure}

\subsection{Ensembled Model Generation from Local Private Models}
The ensembled model construction is the final step of the proposed method. This step is performed by a public cloud when all local private models are received. The set of all differentially private models are represented as: $M_{P} = \{M_{P1}, M_{P1}, \hdots, M_{Pn}\}$, where $n$ is the hospital id. To aggregate the model, a Bootstrap Aggregation or BAGGing \cite{breiman1996bagging} based ensemble averaging technique is used (see Fig. \ref{fig:ensemble}). The averaging function denoted as $\bar{X}$, takes $(p-2)$-th layers output of local private models as input. The operation of $\bar{X}$ can be expressed as:
\begin{equation}\label{eq:avg}
    \bar{X} = \frac{1}{n} \sum_{k=1}^{n} z_{ik},
\end{equation}
where $n$ is the number of private models, and $z_{ik}$ is a specific $z$ value in the $i$th position within an output vector. Next, the results are fed as an input to the last two layers ($L_{p-1}$ and $L_{p}$) of the initial model $M_{I}$. Finally, we get the ensembled model $M_{E}$ which is distributed to all private edge servers.

\begin{algorithm}[ht!]
\SetAlgoNoLine
\caption{Ensemble model construction}
\label{alg: cloudProcess}
\KwIn
{
    \begin{minipage}[t]{10cm}%
     \strut
        Set of local private models, \\$M_{P} = \{M_{P1}, M_{P1}, \hdots, M_{Pn}\}$
        
        Initial model, $M_I$ 
     \strut
  \end{minipage}%
}
\KwOut{
    \begin{minipage}[t]{10cm}%
     \strut
        Ensembled model, $M_E$
     \strut
    \end{minipage}%
}

\textbf{Initialization:}

 Ensembled model, $M_E \leftarrow \emptyset$ \\
 
 Vector of $(p-2)$ layer values $z \leftarrow \emptyset$

    \Foreach{$M_{Pi} \in M_{P}$}{
        z.add($M_{Pi}.getValues(L_{p-2}$))
    }

    $avg \leftarrow computeAvg(z)$ using Eq. (\ref{eq:avg})
    
    $M_{E} \leftarrow generateModel(avg)$
    
    // Add the last two layers for prediction
    
    $M_{E}.addLayer(L_{p-1})$
    
    $M_{E}.addLayer(L_{p})$

\textbf{return} $M_{E}$

\end{algorithm}

%% file: sections/04-experimental-results.tex
\section{Results and Discussion}\label{sec:exp}
In this section, we provide information on the experimental setup used. Then, we perform experiments on MNIST and Lung Cancer datasets to evaluate the model performance of our scheme.

\subsection{Experimental Setup}

\subsubsection{Testing Environment}

We used AWS Sagemaker for our experiment. We chose AWS g4dn.2xlarge machines, which contain 1 NVIDIA T4 GPU with 16 GB GPU memory and 32 GB RAM. The experiments were carried out using Python version 3.7.

\subsubsection{Datasets}
For all experiments, we consider five clients participating in the training process. The training datasets and models are defined as follows:
\begin{itemize}
  \item \textbf{MNIST}. MNIST dataset is a 28 $\times$ 28 multi-class handwritten digits consisting of numbers ranging from 1 to 10. The dataset consists of 70,000 images with training and testing datasets combined. For MNIST dataset experiment, we consider a CNN model as shown in Figure \ref{fig: mnist_cnn}.
  
    \begin{figure}[ht!]
    \centering
    \includegraphics[width=0.9\linewidth]{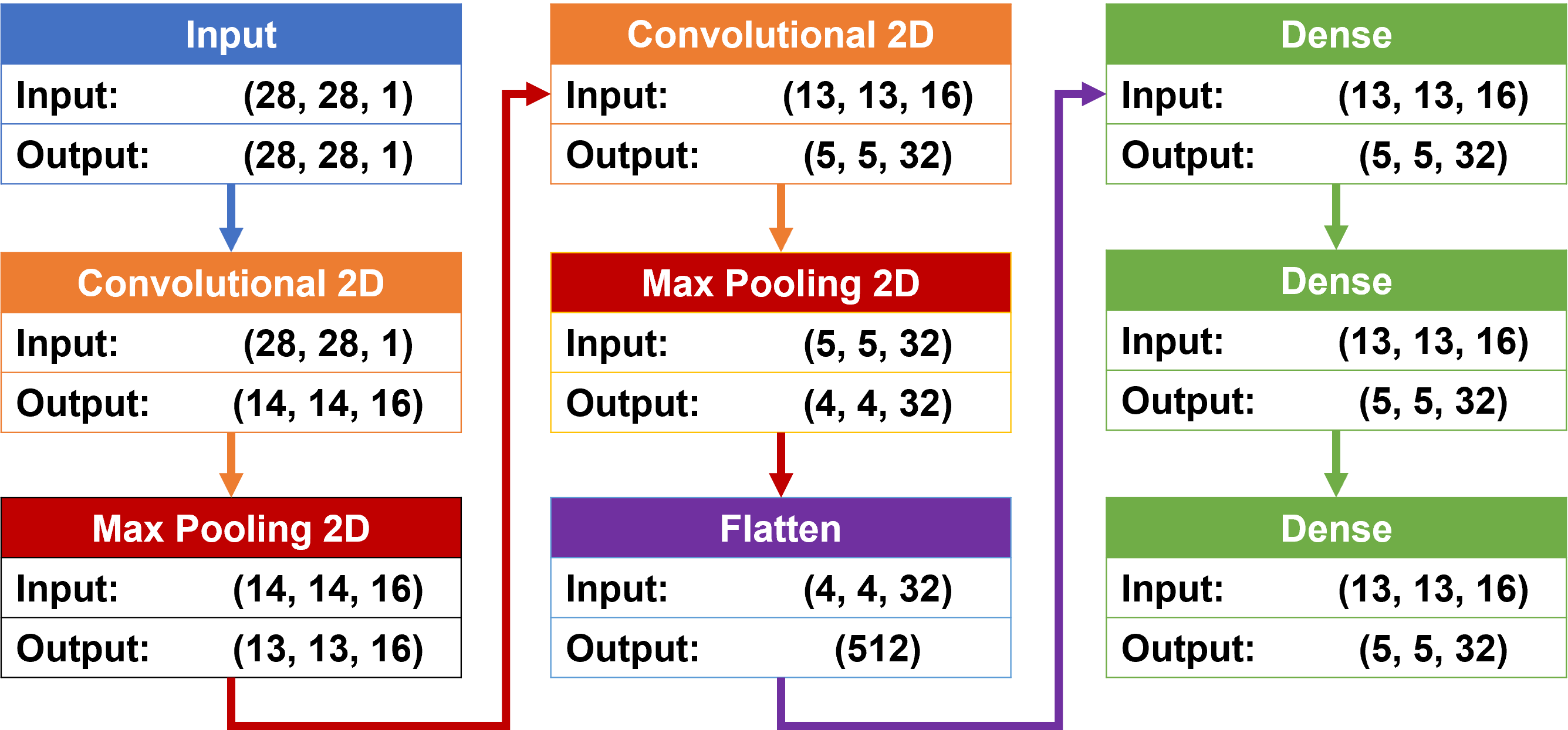}
    \caption{CNN model for MNIST}
    \label{fig: mnist_cnn}
    \end{figure}
  
  \item \textbf{Lung and colon cancer}. Lung and colon cancer dataset retrieved from \cite{borkowski2019lung} consists of 768 $\times$ 768 images from five classes (lung\_n, lung\_scc, lung\_aca, colon\_n,colon\_aca). Each class consists of 5000 images. In this experiment, we are using three classes out of the five classes, which are lung\_n, lung\_scc, and lung\_aca. The CNN model we use for the Lung Cancer dataset is shown in Figure \ref{fig: lnc_cnn}.
  
    \begin{figure}[ht!]
    \centering
    \includegraphics[width=0.9\linewidth]{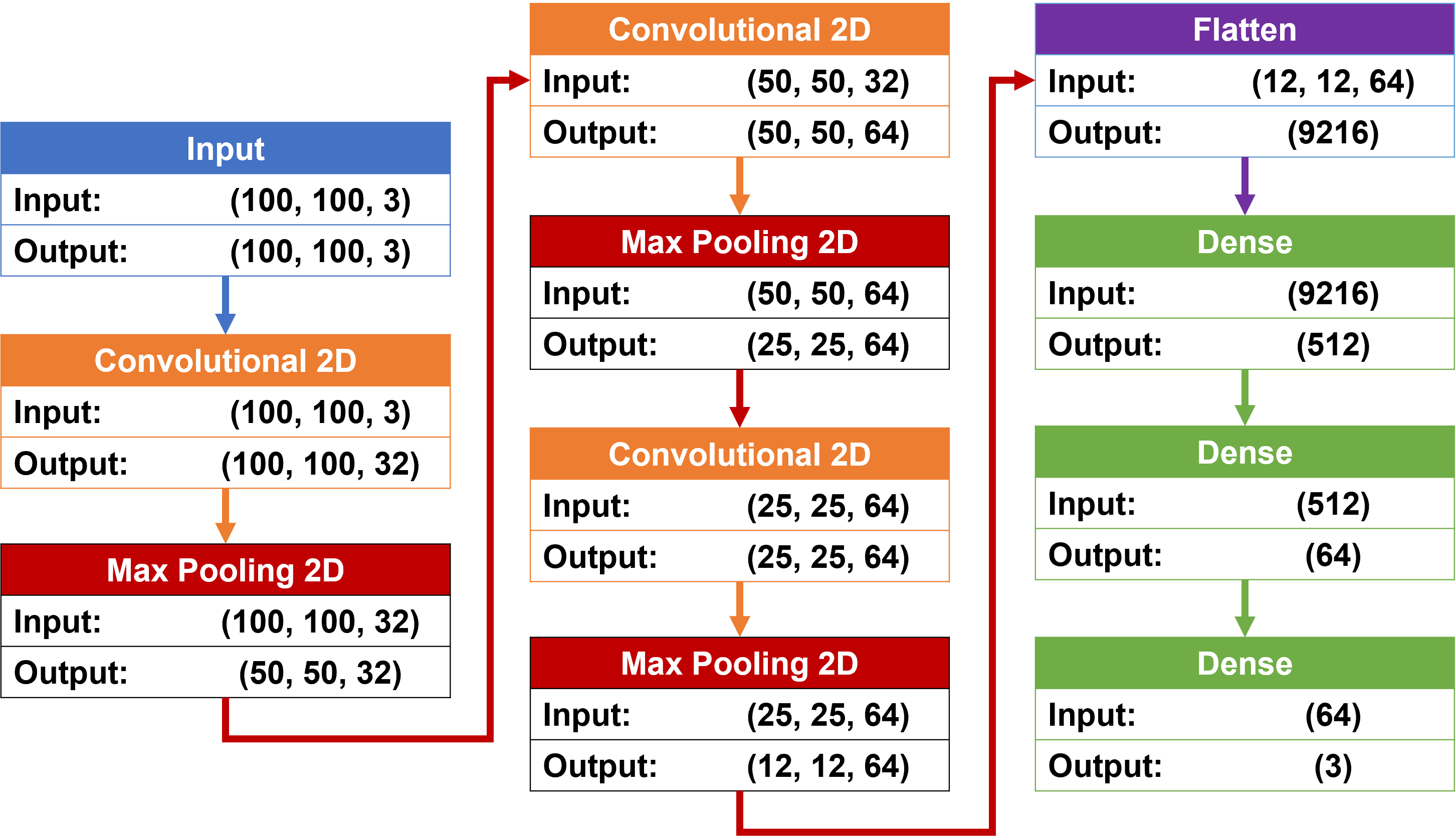}
    \caption{CNN model for Lung Cancer}
    \label{fig: lnc_cnn}
    \end{figure}
   
Data in our experiment is divided into four different parts as shown in Figure \ref{fig: data_part}. We partition our data by assuming a real-world scenario. Validation data is a data partition to represent any unforeseen or future data to be predicted. This partition will be used to test our initial public model and the final ensemble model. Public training data is used to train our initial public model. Its size is set to be smaller than other partitions. Private training data represents the data held by private institutions. This data is used to train our private model. On the other hand, the private test data will be used to test the privately trained model. In order to simplify our experiment, the same data partition sizes will not be changed unless stated otherwise. During our experiments with MNIST dataset, validation data, public data, private training data, and private test data are set to be 28000, 420, 6653, and 1663 respectively. For Lung Cancer dataset, validation data, public data, private train data, and private test data are set to 6000, 90, 1426, and 356, respectively.

 \begin{figure}[ht!]
    \centering
    \includegraphics[width=1\linewidth]{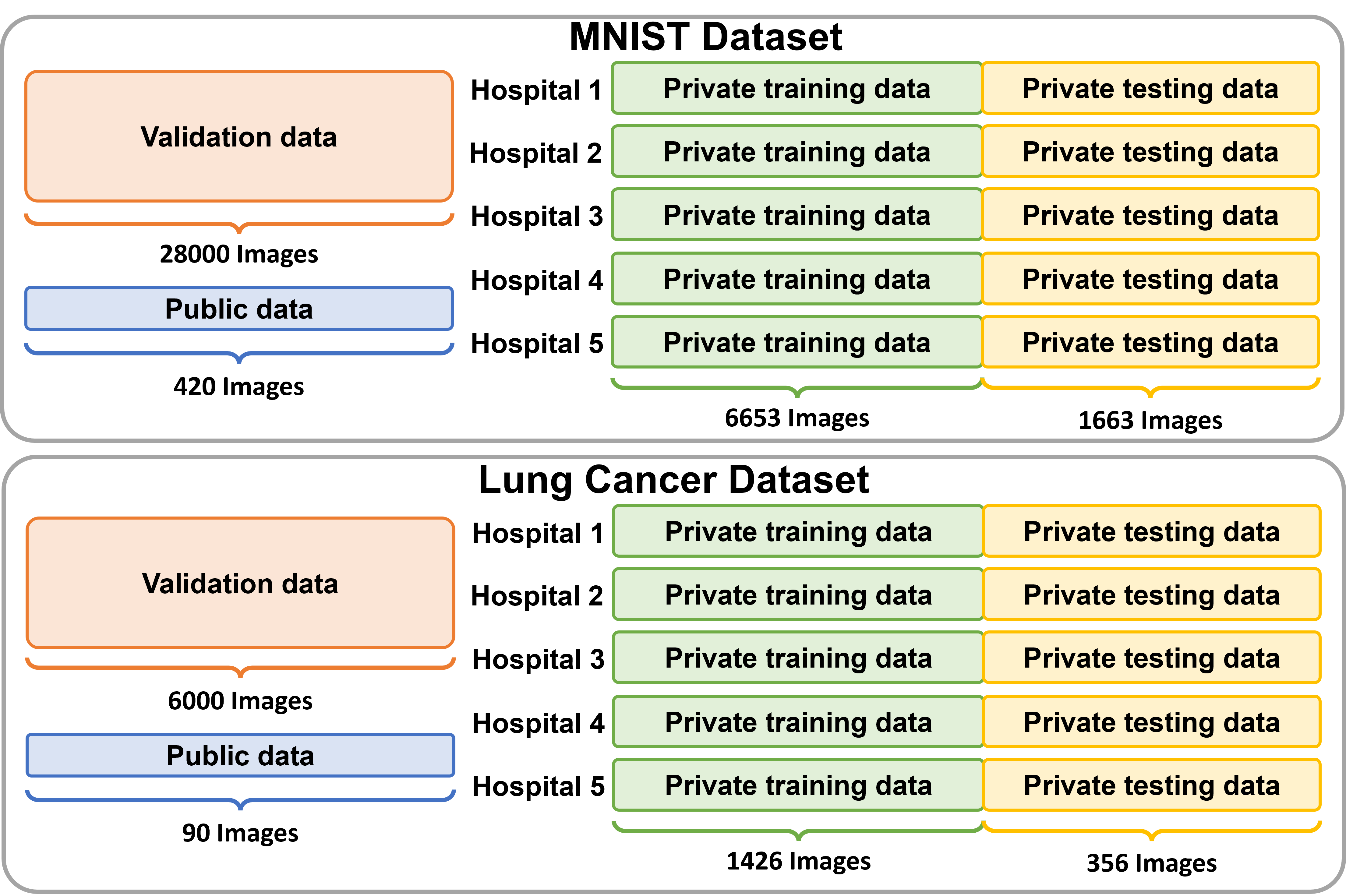}
    \caption{Data partition scheme}
    \label{fig: data_part}
    \end{figure}
  
\end{itemize}

\subsection{Results Analysis}
\begin{figure}[tbh!]
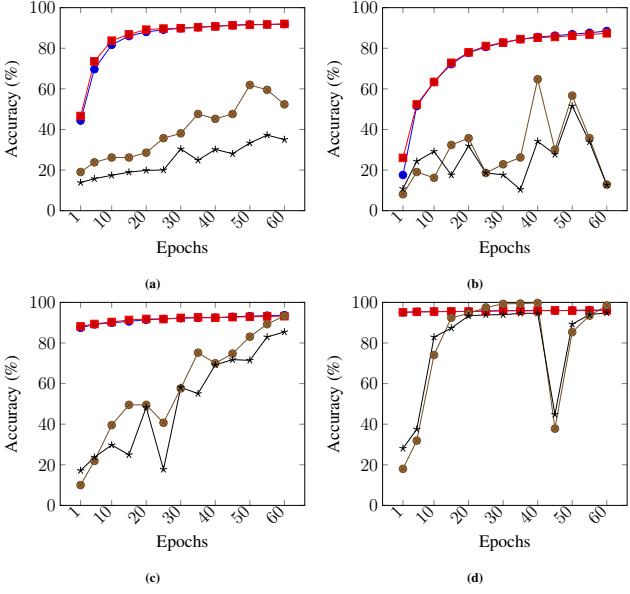

\centering
\begin{subfigure}[tbh!]{0.47\columnwidth}
    \resizebox{1\columnwidth}{!}
    {

        }
    \caption{}
    \label{fig:eff_mnist01}
\end{subfigure}
\begin{subfigure}[tbh!]{0.47\columnwidth}
    \resizebox{1\columnwidth}{!}
    {
        %
        }
    \caption{}
    \label{fig:eff_mnist05}
\end{subfigure}
\begin{subfigure}[tbh!]{0.47\columnwidth}
    \resizebox{1\columnwidth}{!}
    {
        %
        }
    \caption{}
    \label{fig:eff_mnist1}
\end{subfigure}
\begin{subfigure}[tbh!]{0.47\columnwidth}
    \resizebox{1\columnwidth}{!}
    {
        %
        }
    \caption{}
    \label{fig:eff_mnist5}
\end{subfigure}
\caption{The effect of public data size used to train the initial public model towards the private model accuracy on MNIST data. The plot shows the accuracies of private model training \ref{privTrainAcc}, private model testing \ref{privTestAcc}, public model training \ref{pubTrainAcc}, and public model testing \ref{pubTestAcc}}
\label{fig:mnistEff}
\end{figure}

\begin{figure}[tbh!]
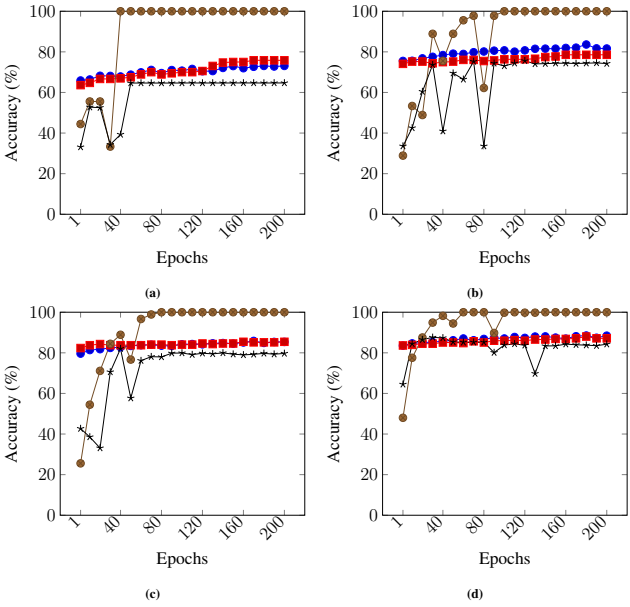

\centering
\begin{subfigure}[tbh!]{0.47\columnwidth}
    \resizebox{1\columnwidth}{!}
    {
        %
        }
    \caption{}
    \label{fig:eff_lnc01}
\end{subfigure}
\begin{subfigure}[tbh!]{0.47\columnwidth}
    \resizebox{1\columnwidth}{!}
    {
        %
        }
    \caption{}
    \label{fig:eff_lnc05}
\end{subfigure}
\begin{subfigure}[tbh!]{0.47\columnwidth}
    \resizebox{1\columnwidth}{!}
    {
        %
        }
    \caption{}
    \label{fig:eff_lnc1}
\end{subfigure}
\begin{subfigure}[tbh!]{0.47\columnwidth}
    \resizebox{1\columnwidth}{!}
    {
        %
        }
    \caption{}
    \label{fig:eff_lnc5}
\end{subfigure}
\caption{The effect of public data size used to train the initial public model towards the private model accuracy on Lung Cancer data. The plot shows the accuracies of private model training \ref{privTrainAcc}, private model testing \ref{privTestAcc}, public model training \ref{pubTrainAcc}, and public model testing \ref{pubTestAcc}}
\label{fig:lncEff}
\end{figure}

For our experiment of results analysis, we first observe the effect of public data size used to train the initial public model towards the private model training accuracy. In Fig. \ref{fig:mnistEff} and Fig. \ref{fig:lncEff}, we use different number of public data size to train our initial public model.

Figure \ref{fig:mnistEff} shows the experiments using the MNIST dataset. we use four settings on the public and private datasets. The size of public and private data used for training is defined as follows: (a): 42 public data and 6713 private data; (b): 210 public data and 6687 private data; (c): 420 public data and 6653 private data; (d): 2100 public data and 6384 private data. For the model training configuration setup, the MNIST dataset is trained in 60 epochs, and the batch size is set to 250. In regards to privacy preservation parameter, we set the clipping threshold value to 1.5

Figure \ref{fig:lncEff} shows the experiments for Lung Cancer dataset. We also used four different settings. The size of public and private data used for training is defined as follows: (a): 9 public data and 1439 private data; (b): 45 public data and 1433 private data; (c): 90 public data and 1426 private data; (d): 450 public data and 1368 private data. For the model training configuration setup, the Lung Cancer dataset is trained in 200 epochs, and the batch size is set to 18. The clipping threshold value used for Lung Cancer dataset experiment is 1.0

As can be seen in both figures, the increase of public data size increases private model accuracy. However, the growth of private model accuracy becomes less significant as the initial public model performance increases. 

\begin{figure}[tbh!]
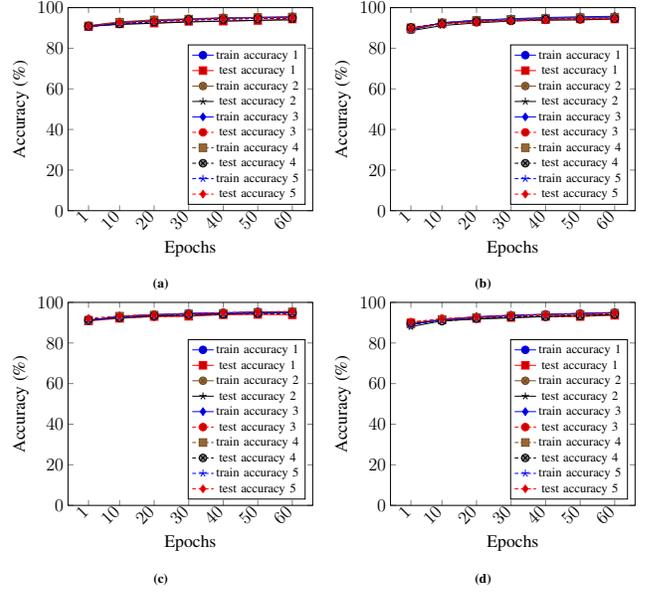

\centering
\begin{subfigure}[tbh!]{0.47\columnwidth}
    \resizebox{1\columnwidth}{!}
    {

        }
    \caption{}
    \label{fig:mnist0.9}
\end{subfigure}
\begin{subfigure}[tbh!]{0.47\columnwidth}
    \resizebox{1\columnwidth}{!}
    {
        %
        }
    \caption{}
    \label{fig:mnist1.1}
\end{subfigure}
\begin{subfigure}[tbh!]{0.47\columnwidth}
    \resizebox{1\columnwidth}{!}
    {
        %
        }
    \caption{}
    \label{fig:mnist1.3}
\end{subfigure}
\begin{subfigure}[tbh!]{0.47\columnwidth}
    \resizebox{1\columnwidth}{!}
    {
        %
        }
    \caption{}
    \label{fig:mnist1.5}
\end{subfigure}
\caption{Private model training and testing accuracies on MNIST dataset with respect to noise multiplier. (a) with noise multiplier 0.9; (b) with noise multiplier 1.1; (c) with noise multiplier 1.3; (d) with noise multiplier 1.5}
\label{fig:mnistAcc}
\end{figure}

\begin{figure}[tbh!]
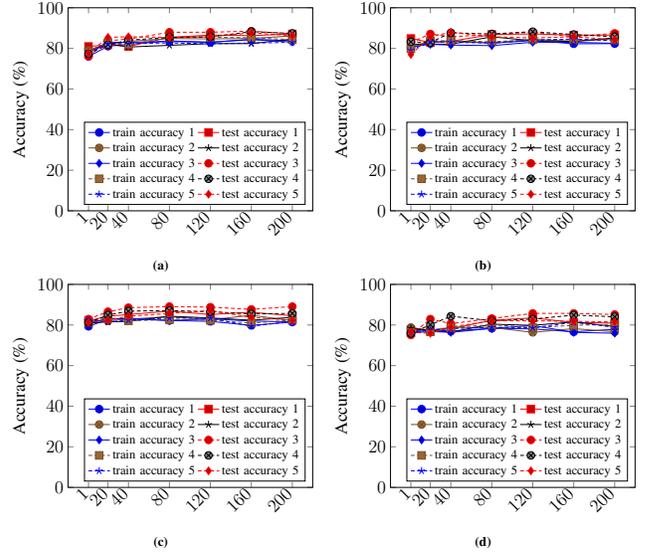

\centering
\begin{subfigure}[tbh!]{0.47\columnwidth}
    \resizebox{1\columnwidth}{!}
    {
        %
        }
    \caption{}
    \label{fig:lnc0.9}
\end{subfigure}
\begin{subfigure}[tbh!]{0.47\columnwidth}
    \resizebox{1\columnwidth}{!}
    {
        %
        }
    \caption{}
    \label{fig:lnc1.1}
\end{subfigure}
\begin{subfigure}[tbh!]{0.47\columnwidth}
    \resizebox{1\columnwidth}{!}
    {
        %
         }
    \caption{}
    \label{fig:lnc1.3}
\end{subfigure}
\begin{subfigure}[tbh!]{0.47\columnwidth}
    \resizebox{1\columnwidth}{!}
    {
        %
        }
    \caption{}
    \label{fig:lnc1.5}
\end{subfigure}
\caption{Private model training and testing accuracies on Lung Cancer dataset with respect to noise multiplier. (a) with noise multiplier 0.9; (b) with noise multiplier 1.1; (c) with noise multiplier 1.3; (d) with noise multiplier 1.5}
\label{fig:lncAcc}
\end{figure}

Next, we tried to observe the effect of the noise multiplier on the private model training from the two datasets. Fig. \ref{fig:mnistAcc} shows the experiment of using 0.9, 1.1, 1.3, and 1.5 noise multipliers on MNIST dataset. Results show that there is no significant difference between the four cases. We employ the same configuration for its noise multiplier for Lung Cancer dataset. Compared to MNIST dataset, the increase of noise in Lung Cancer dataset results in dispersed accuracy on each private model (see Fig. \ref{fig:lncAcc}).

We summarise our experiments in Table \ref{table: mnist} and Table \ref{table: lnc}. Table \ref{table: mnist} exhibits a summary of model performance on MNIST dataset when the initial model is trained with 0.001 learning rate and private model with 0.15 learning rate. Table \ref{table: lnc} presents the summary of model performance on Lung Cancer dataset when the initial model is trained with 0.001 learning rate and private model with 0.015 learning rate.

From Table \ref{table: mnist}, it can be seen that the final ensemble model accuracy tends to provide accuracy higher than the average of all private models accuracies. However, when the $\epsilon$ value is 1.9, the ensemble model accuracy remains the same as the average of all private models' accuracy.

\begin{table}[!th]
\renewcommand{\arraystretch}{1.3}
\caption{Model performance and privacy configuration on MNIST dataset}
\label{table: mnist}
\centering
\scalebox{0.8}{
\begin{tabular}{|c|c|c|c|c|c|c|c|c|}
\hline
\makecell{ Noise \\ multiplier } & \makecell{ Clipping \\ threshold } & $\delta$ & $\epsilon$ & \makecell{ Initial \\ model \\ accuracy \\ (\%)} &\makecell{ Private \\ models \\ average \\ accuracy \\ (\%)}& \makecell{ Final \\ model \\ accuracy \\ (\%)}\\
\hline
$0.9$ & $1.0$ & $1e-4$ & $12$ & $87.2\%$ & $94.6\%$ & $94.9\%$\\
\hline
$1.1$ & $1.0$ & $1e-4$ & $8$ & $85\%$ & $93.2\%$ & $93.7\%$\\
\hline
$1.3$ & $1.0$ & $1e-4$ & $6$ & $87.2\%$ & $93.9\%$ & $94.3\%$\\
\hline
$1.5$ & $1.0$ & $1e-4$ & $4.8$ & $90.8\%$ & $94.6\%$ & $94.8\%$\\
\hline
$3$ & $1.0$ & $1e-4$ & $1.9$ & $81.8\%$ & $93\%$ & $93\%$\\
\hline
\end{tabular}
}
\end{table}

While slight improvement can be seen on the final ensembled model in Table \ref{table: mnist}, a more significant accuracy improvement on the final ensemble model can be seen on the Lung Cancer dataset (see Table \ref{table: lnc}), specifically when the value of $\epsilon$ is 1.7, improvement of final model accuracy reaches up to 10 percent from the private model accuracy. However, despite having a significant improvement on final model accuracy, when the noise is big enough, the performance of the private model lies below the initial model accuracy.

\begin{table}[!th]
\renewcommand{\arraystretch}{1.3}
\caption{Model performance and privacy configuration on Lung Cancer dataset}
\label{table: lnc}
\centering
\scalebox{0.8}{
\begin{tabular}{|c|c|c|c|c|c|c|c|c|c|}
\hline
\makecell{ Noise \\ multiplier } & \makecell{ Clipping \\ threshold } & Delta & $\epsilon$ & \makecell{ Initial \\ model \\ accuracy \\ (\%)} &\makecell{ Private \\ models \\ average \\ accuracy \\ (\%)}& \makecell{ Final \\ model \\ accuracy \\ (\%)}\\
\hline
$0.9$ & $1.0$ & $7e-4$ & $11.5$ & $79.1\%$ & $87.8\%$ & $88.7\%$\\
\hline
$1.1$ & $1.0$ & $7e-4$ & $7.6$ & $78.6\%$ & $85.3\%$ & $88.4\%$\\
\hline
$1.3$ & $1.0$ & $7e-4$ & $5.6$ & $77\%$ & $83.5\%$ & $87.5\%$\\
\hline
$1.5$ & $1.0$ & $7e-4$ & $4.5$ & $78\%$ & $83.8\%$ & $87.6\%$\\
\hline
$3$ & $1.0$ & $7e-4$ & $1.7$ & $75.3\%$ & $68.8\%$ & $79.9\%$\\
\hline
\end{tabular}
}
\end{table}

Finally, we compare our proposed method's performance with the existing methods, TrPATE \cite{8689016} and COFEL \cite{lian2021cofel} on MNIST dataset to prove the effectiveness of our strategy. For comparison, we use the same model configurations provided in the paper. The results are summarised in Table \ref{table: compare}. As can be seen from the table, our method outperformed the other existing method even for a smaller value of $\epsilon$.

\begin{table}[!th]
\renewcommand{\arraystretch}{1.3}
\caption{Model performance comparison}
\label{table: compare}
\centering
\scalebox{0.8}{
\begin{tabular}{|c|c|c|c|c|c|}
\hline
\multicolumn{2}{|c|}{TrPATE} & \multicolumn{2}{c}{COFEL} & \multicolumn{2}{|c|}{Proposed Method} \\ 
\hline
$\epsilon$ & $accuracy (\%)$ & $\epsilon$ & $accuracy (\%)$ & $\epsilon$ & $accuracy (\%)$\\
\hline
$1.3$ & $90\%$ & $1$ & $88\%$ & $1.9$ & $92.8\%$\\
\hline
$5.8$ & $93\%$ & $5$ & $90\%$ & $4.8$ & $94.8\%$ \\
\hline
$8.8$ & $93\%$ & $10$ & $91.8\%$ & $8$ & $94.4\%$\\
\hline
$15.7$ & $93.2\%$ & $15$ & $92.4\%$ & $12$ & $94.9\%$\\
\hline
\end{tabular}
}
\end{table}

%% file: sections/05-conclusion.tex
\section{Conclusion}\label{sec:con}
This paper proposes an ensemble and transfer learning infused framework for privacy-preserving DNN model generation in IoT, edge, and cloud convergence. Differential Privacy is used to add noise in the local model to ensure privacy of the model. As adding noise to the model significantly reduces the model performance, Transfer Learning is used with CNN to reduce the loss and improve the efficiency of the local model. The proposed framework involves multiple participants. Hence, local models from different participants are ensembled at the cloud to generate a collective learning model, called the final model, to ensure higher prediction accuracy. 
From our experiment, we demonstrated that transferring the knowledge of public data in ensemble learning enhances the accuracy of the final model. The effectiveness of our model has been compared against state-of-the-art methods such as TrPATE and COFEL. Experimental results show that our method outperformed the existing work. 
In this article, we assume that all participants use the same knowledge domain. Further research should also investigate the performance while transferring knowledge from different domains.

\section*{Acknowledgement}
This work is supported by the Australian Research Council Discovery Project (DP210102761).